\documentclass[conference,twocolumn,10pt]{IEEEtran}
\usepackage{graphicx}
\usepackage{amsmath,amssymb}
\usepackage{mathtools}
\usepackage{dsfont}
\usepackage{cases}
\usepackage{bm}
\usepackage{amsfonts}
\usepackage[font=footnotesize]{caption}
\usepackage{indentfirst}
\usepackage{color}
\usepackage{cite}
\usepackage{caption}
\usepackage{algorithm}
\usepackage{algpseudocode}
\usepackage{booktabs}
\usepackage{subfigure}
\usepackage{multirow}
\usepackage{amsthm}
\usepackage{setspace}

\theoremstyle{remark}
\newtheorem{remark}{Remark}
\theoremstyle{}
\newtheorem{theorem}{Theorem}
\theoremstyle{}

\newtheorem{lemma}{Lemma}

\theoremstyle{}
\newtheorem{definition}{Definition}
\theoremstyle{remark}

\theoremstyle{definition}

\makeatletter
\newcommand{\tabcaption}{\def\@captype{table}\caption}
\makeatletter
 \allowdisplaybreaks[4]

\ifCLASSINFOpdf
\else
\fi

\begin{document}

\title{Storage, Computation, and Communication: A Fundamental  Tradeoff in  Distributed Computing }
\author{\IEEEauthorblockN{Qifa Yan}
\IEEEauthorblockA{ LTCI, T\'el\'ecom ParisTech\\ %\\Universite Paris-Saclay,\\
75013 Paris, France\\
Email: qifa.yan@telecom-paristech.fr}
\and
\IEEEauthorblockN{Sheng Yang\\   }
\IEEEauthorblockA{L2S, CentraleSup\'elec\\
%3 rue Joliot-Curie,\\
91190 Gif-sur-Yvette, France\\
Email:sheng.yang@centralesupelec.fr}
\and
\IEEEauthorblockN{Mich\`ele Wigger}
\IEEEauthorblockA{LTCI, T\'el\'ecom ParisTech \\
%Universite Paris-Saclay,\\
75013 Paris, France\\
Email:  michele.wigger@telecom-paristech.fr}
}

%\author{Michael~Shell,~\IEEEmembership{Member,~IEEE,}
%        John~Doe,~\IEEEmembership{Fellow,~OSA,}
%        and~Jane~Doe,~\IEEEmembership{Life~Fellow,~IEEE}% <-this % stops a space
%\thanks{M. Shell is with the Department
%of Electrical and Computer Engineering, Georgia Institute of Technology, Atlanta,
%\thanks{Manuscript received April 19, 2005; revised December 27, 2012.}}

\maketitle

\IEEEpeerreviewmaketitle
%\begin{spacing}{2.0}
\begin{abstract} %Distributed computing has become one of the most important frameworks in dealing with large computation tasks. In this paper,
  We consider a MapReduce-like distributed computing system.
  We derive a lower bound on the  communication cost for any given storage  and computation costs. This lower bound matches the achievable bound we proposed recently. As a result,
  % the optimal tradeoff between storage space, computation and communication loads is characterized,  i.e., a
  we completely characterize the optimal tradeoff between the
  storage, the computation, and the communication. Our
  result generalizes the previous one by Li \emph{et al.} to also account for the number of computed intermediate values.
 %Different from the coded distributed computing (CDC) framework introduced by Li \emph{et al.},  we measure the computation load by  normalizing the number of actually computed intermediate values by the number of total intermediated values, instead of the number of totally stored files normalized by the number of total files.
  % In particular, the optimal communication load for a given total cache size constraint is in accordance with that in the CDC, while the computation load required to achieve the optimal communication load is significant lower than that in CDC scheme in our framework.

\end{abstract}

\section{Introduction}

Systems like  MapReduce \cite{MapReduce}, Dryad  \cite{Dryad} \emph{etc.} have become popular platforms  for distributed computing to perform data-parallel computations across  distributed computing nodes. In such systems, the  computations  are typically decomposed into ``Map" and ``Reduce" functions as detailed in the following.
Consider the task of computing $K$ output functions of the form
  \begin{IEEEeqnarray}{rCl}
  	\phi_k(w_1,\cdots,w_N)&=&h_k(g_{k,1}(w_1),\cdots,g_{k,N}(w_N)),\IEEEeqnarraynumspace\label{eqn:hk}\\
  	&&\qquad\qquad\qquad k=1,\cdots,K.\notag
  \end{IEEEeqnarray}
Here, each output function $\phi_k$  depends on all $N$  data blocks $w_1,\cdots,w_N$, but can be decomposed into:
\begin{itemize}
	\item $N$ \emph{map functions} $g_{k,1},\cdots,g_{k,N}$, each only depending  on one block;
	\item  a \emph{reduce function} $h_k$ that combines the outcomes of the $N$ map functions.
\end{itemize}

%, carried out through three main steps:
%Firstly,  each node stores a subset of the input data  locally, and calculates some intermediate values (IVAs) according to the designed map functions. Next,  the distributed nodes exchange the IVAs computed, so that each node is able to  collect  its desirable IVAs. Finally,  each node calculates  out its allocated  output functions, based on the IVAs computed in the Map phase and collected in the Shuffling phase.

%On the other hand, in caching networks, a technique named \emph{coded caching}  originally proposed  by Maddah-Ali and Niesen to explore the \emph{multicast} opportunities to reduce the communication load has been received significant attention in the literature \cite{Maddah2014Fundamental,Maddah2015Decentralized}. It   has been investigated in various settings such as device-to-device wireless networks \cite{Ji2016D2D}, online file update \cite{Pedarsani2016online,Yan2017online}, multiple antennas \cite{Ngo2018Scalable,Caire2017multiantenna}, Gaussian broadcast channels \cite{Bidokhti2017Gaussian}, placement delivery arrays \cite{Yan2017PDA,Yan2018bipartite} \emph{et. al.}.

Computation of such functions can be performed in a distributed way following  3-phases: In the first \emph{map phase},  each node locally stores a subset of the input data $\mathcal{M}_k\subseteq\{w_1,\cdots,w_N\}$, and calculates all \emph{intermediate values} (IVAs) that  depend on the stored data:
\begin{IEEEeqnarray}{c}
	\{g_{l,n}(w_n):l\in\{1,\cdots,K\},~w_n\in\mathcal{M}_k\}.\notag
\end{IEEEeqnarray}
%The storage cost  $r$, is measured by  the total number of files stored across the nodes normalized by the number of files.
In the subsequent \emph{shuffle phase}, the nodes exchange the IVAs computed during the map phase, so that each node $k$ is aware of all the IVAs $g_{k,1}(w_1),\cdots,g_{k,N}(w_N)$ required to calculate its own output function $\phi_k$. % The communication cost $L$, is measured by the total amount of information exchanged in the shuffle phase normalized by the total size of  all $KN$ IVAs.
In the final \emph{reduce phase}, each node~$k$ combines the IVAs  with the reduce function $h_k$ as indicated in \eqref{eqn:hk}.

Li \emph{et al.} \cite{Li2018Tradeoff} proposed a  scheme, termed \emph{coded distributed computing} (CDC), that in the map phase stores files multiple times across users so as to enable multicast opportunities for the shuffle phase. This approach can significantly reduce the communication load over traditional schemes, and was proved in \cite{Li2018Tradeoff} to have the smallest communication load among all the distributed computing schemes with  same total storage requirements.  %Subsequently, various works extended the results in \cite{Li2018Tradeoff}. For example,    straggling nodes are investigated  in \cite{Lee2018codes,Li2016unified}; \cite{Qian2017How}  studies  optimal allocation of computation resources; \cite{Li2017Framework} \emph{et. al.} extends the works to wireless networks.
Some extensions have been made in follow-up works. For example,
straggling nodes were investigated  in \cite{Lee2018codes};
\cite{Qian2017How}  studied optimal allocation of computation
resources; \cite{Li2017Framework} considered distributed
nodes in a wireless network.

It is worth mentioning that
Li \emph{et al.} in \cite{Li2018Tradeoff} used the term
\emph{computation-communication} tradeoff, because they assumed that
each node calculates all the IVAs that can be obtained from the
data stored at that node, irrespective of whether these IVAs are used in the
sequel or not.
In this sense, the total number of calculated IVAs is
actually a measure of the total storage space consumed across the nodes.
This is why we would rather refer to it as the \emph{storage-communication} tradeoff.

Naturally,  if an IVA is not used subsequently, there is no need to compute it, which can save  computation resources (e.g., power) and shorten  calculation latency. Therefore,
 it is natural to
investigate a more general framework, where each node is allowed to choose to calculate or not the IVA for each output function from the data stored locally.
%In the CDC scheme, intuitively, there exists some redundancy in computing, since  it was assumed that each node calculate IVAs for all output functions from all the files it stored. The computation load was thus   measured by the number of files mapped or stored by each node normalized by the number of total files.  To remove the redundancy in computing, it is suitable to measure the computation load in a finer way: enumerate the number of \emph{actually computed} IVAs and normalize it by the total number of IVAs.
The number of IVAs that each node needs to calculate normalized by the
total number of IVAs is then used to measure the real computation load.
In this sense, we extend the storage-communication tradeoff  in
\cite{Li2018Tradeoff} to a \emph{storage-computation-communication}
tradeoff. In particular, we wish to  characterize the smallest
communication load required in the shuffle phase for a given storage
space and a given number of IVAs  calculated during the map phase.
  Ezzeldin \emph{et al.} proposed a modification on the CDC scheme in \cite{Fragouli}, that compute IVAs only if they are used subsequently.
Recently, we also proposed a new scheme named \emph{distributed computing
and coded communication (D3C)}  \cite{Yan2018SCC}, and derived the
tradeoff achieved by this scheme. In this paper, we provide a matching converse,
and thereby characterize completely the optimal
storage-computation-communication tradeoff.

%The  paper is organized as follows:  Section \ref{sec:model} introduces the system model, and Section \ref{sec:result} presents  the main result. Then we present the converse proof in Section  \ref{sec:converse} and conclude this paper in Section \ref{sec:conclusion}.

\emph{Notations:} Let $\mathbb{N}^+$ denote the set of positive integers, and for $m,n\in\mathbb{N}^+$, let  $\mathbb{F}_{2^m}^n$ denote the $n$-dimensional vector space over the finite field $\mathbb{F}_{2^m}$. We also abbreviate  $\{1,\cdots,n\}$ by $[n]$. For scalar quantities we use (upper or lower case) standard font, for sets calligraphic font, and for collections (sets of sets) bold font.  The cardinality of a set $\mathcal{A}$ is denoted  $|\mathcal{A}|$. The indicator function of an event is written as $\mathbb{I}(\cdot)$.

\section{System Model}\label{sec:model}

Consider a system with $K$ distributed computing nodes and $N$ files.
Specifically, given any $N$ files
\begin{IEEEeqnarray}{c}
\mathcal{W}=\{w_1,\cdots,w_N\},\quad w_i\in \mathbb{F}_{2^F},\forall~i\in[N].\notag
 \end{IEEEeqnarray}
 Node $k$ ($k\in[K]$) wishes to compute an output function $\phi_k:\mathbb{F}_{2^{F}}^N\rightarrow \mathbb{F}_{2^B}$, which maps all the files to a bit stream $u_k=\phi_k(w_1,\cdots,w_N)\in\mathbb{F}_{2^B}$ of length $B$, where $B\in\mathbb{N}^+$.

Following the MapReduce framework \cite{Li2018Tradeoff,Li2017Framework},
we assume that the computation of the output functions $\phi_k$ can be
decomposed as in \eqref{eqn:hk},
%\begin{IEEEeqnarray}{c}
%\phi_k(w_1,\cdots,w_N)=h_k(g_{k,1}(w_1),\cdots,g_{k,N}(w_N)),
%\end{IEEEeqnarray}
where
\begin{itemize}
  \item The ``Map" function
\begin{IEEEeqnarray}{c}
g_{k,n}: \mathbb{F}_{2^F}\rightarrow \mathbb{F}_{2^T},~k\in[K],~ n\in[N]\notag
\end{IEEEeqnarray}
maps the file $w_n$ into a binary intermediate value (IVA) of length $T$, i.e., $v_{k,n}\triangleq g_{k,n}(w_n)\in \mathbb{F}_{2^T}$, where  $T\in\mathbb{N}$.
  \item The ``Reduce" function
\begin{IEEEeqnarray}{c}
  h_k: \mathbb{F}_{2^T}^N\rightarrow \mathbb{F}_{2^B}, ~k\in[K]\notag
\end{IEEEeqnarray}
  maps the intermediate values
  \begin{IEEEeqnarray}{c}
  \mathcal{V}_k\triangleq\{v_{k,n}:n\in[N]\}\notag
  \end{IEEEeqnarray}
    into the output stream $u_k=h_k(v_{k,1},\cdots,v_{k,N})$.
\end{itemize}

%A trivial decomposition  is setting the map functions be identity functions and the reduce functions be the output functions, i.e., $g_{k,n}(w_n)=w_n$ and $h_k=\phi_k$, $\forall~n\in[N],~k\in[K]$. Thus, the decomposition always exists.

    The computations are carried out in three phases.

  1) \textbf{Map Phase:} Each node $k$ stores a subset of files
  $\mathcal{M}_k\subseteq \mathcal{W}$, $k\in[K]$, and then for each
  file $w_n\in\mathcal{M}_k$,  computes a subset of IVAs
  $\mathcal{C}_{k,n}=\{v_{q,n}:q\in \Lambda_{k,n}\}$, where
  $\Lambda_{k,n}\subseteq[K]$. Denote the set of IVAs computed at node $k$ by $\mathcal{C}_k$, i.e.,
\begin{IEEEeqnarray}{c}
    \mathcal{C}_k\triangleq \bigcup_{n:w_n\in\mathcal{M}_k}\mathcal{C}_{k,n}.\label{eqn:Ck}
\end{IEEEeqnarray}

To measure the storage and computation  cost of the system, we introduce
the following two definitions.

\begin{definition}[Storage Space]
  We define the \emph{storage space} $r$, as the total number of files stored across the $K$ nodes, normalized by the total number of files $N$, i.e.,
\begin{IEEEeqnarray}{c}
r\triangleq\frac{\sum_{k=1}^K|\mathcal{M}_k|}{N}.\label{eqn:r}
\end{IEEEeqnarray}
\end{definition}

       \begin{definition}[Computation Load] We define the computation load $c$, as the total number of map functions computed across the  $K$ nodes,  normalized by the total number of map functions $NK$, i.e.,
  \begin{IEEEeqnarray}{c}
       c\triangleq\frac{\sum_{k=1}^K|\mathcal{C}_k|}{NK}.\label{eqn:c}
\end{IEEEeqnarray}
       \end{definition}

 2) \textbf{Shuffle Phase:} To compute the output function $\phi_k$,
 node $k$ needs to collect the IVAs of $\phi_k$ that are not computed
 locally in the map phase, i.e., $\mathcal{V}_k\backslash
 \mathcal{C}_k$. After the map phase, the  $K$ nodes  exchange the
 computed IVAs. Particularly, each node $k$ creates and multicasts a
 signal $X_k\in\mathbb{F}_{2^{l_k}}$ for some $l_k\in\mathbb{N}$, as a
 function of the IVAs computed in the map phase, namely,
\begin{IEEEeqnarray}{c}
       X_k=\varphi_k\left(\mathcal{C}_k\right)\notag
\end{IEEEeqnarray}
        to all the other nodes for some encoding function
 \begin{IEEEeqnarray}{c}
       \varphi_k: \mathbb{F}_{2^T}^{|\mathcal{C}_k|}\rightarrow \mathbb{F}_{2^{l_k}}.\notag
\end{IEEEeqnarray}
      All the nodes receive the signals $X_1,\cdots,X_K$ error-free.

        \begin{definition}[Communication Load] We define the
          communication load $L$, as the total number of the bits transmitted by the $K$ nodes during the shuffle phase normalized by the total length of all intermediate values $NKT$, i.e.,
\begin{IEEEeqnarray}{c}
        L\triangleq\frac{\sum_{k=1}^K l_k}{NKT}.\notag
\end{IEEEeqnarray}
        \end{definition}

3) \textbf{Reduce Phase:}   With the  signals $\{X_i\}_{i=1}^K$ exchanged during the shuffle phase and the IVAs $\mathcal{C}_k$ computed  locally  in map phase,
       node $k$  restores all the IVAs in $\mathcal{V}_k$, i.e.,
  \begin{IEEEeqnarray}{c}
       (v_{k,1},\cdots,v_{k,N})=\psi_k\left(X_1,\cdots,X_K,\mathcal{C}_k\right),\notag
\end{IEEEeqnarray}
with the function
  \begin{IEEEeqnarray}{c}
       \psi_k: \mathbb{F}_{2^{l_1}}\times\mathbb{F}_{2^{l_2}}\times\cdots\mathbb{F}_{2^{l_K}}\times \mathbb{F}_{2^T}^{|\mathcal{C}_k|}\rightarrow \mathbb{F}_{2^{T}}^{N}.\notag
\end{IEEEeqnarray}
     Finally, it proceeds to compute
 \begin{IEEEeqnarray}{c}
       u_k=h_k(v_{k,1},\cdots,v_{k,N}).\notag
\end{IEEEeqnarray}

\begin{definition}\label{definition:ach}
 A distributed computing system is said to achieve
a storage-computation-communication (SCC) triple $(r, c, L)$, if for any $\epsilon>0$, when $N$
is sufficiently large,
there exists a map-shuffle-reduce procedure such that  the storage space, computation
load, and communication load do not exceed $r+\epsilon$, $c+\epsilon$, and $L+\epsilon$, respectively.
In particular, we define the optimal communication
load by
\begin{IEEEeqnarray}{c}
L^*(r,c)\triangleq\inf\big\{L:(r,c,L)~\mbox{is achievable}\big\}.\notag
\end{IEEEeqnarray}
\end{definition}

Without loss of generality (W.L.O.G), we assume
$
1\leq c\leq r<K
$.
In fact,  $|\mathcal{C}_k|\leq |\mathcal{M}_k|K$ is implied by \eqref{eqn:Ck}, and   thus $c\leq r$ by \eqref{eqn:r} and \eqref{eqn:c}. Moreover, since each IVA needs to be computed at least once somewhere, we have $c\geq 1$. Furthermore, if $r\geq K$, each node  trivially stores all the files and locally computes all the IVAs required for its output function.

%\begin{remark}\label{remark1} If each node $k$ computes the IVAs required by all the output functions from all the input files in its caches,  i.e., $\mathcal{C}_k=\{v_{k,n}:n\in\mathcal{M}_k\}$,   then $|\mathcal{C}_k|=|\mathcal{M}_k|\cdot K$, thus the computation load is given by
%\begin{IEEEeqnarray}{c}
%c=\frac{\sum_{k=1}^K|\mathcal{C}_k|}{NK}=\frac{\sum_{k=1}^K|\mathcal{M}_k|\cdot K}{NK}=r.
%\end{IEEEeqnarray}
%This is exactly the case  investigated in \cite{Li2018Tradeoff}, in  which the authors establishes the fundamental storage-communication  tradeoff for this case. So, when $c=r$,  the storage-computation-communication tradeoff is \cite{Li2018Tradeoff}
%\begin{IEEEeqnarray}{c}
%L^*(r,r)=\mbox{Lcov}
%\left\{\left(r,\frac{1}{r}\left(1-\frac{r}{K}\right)\right):r=1,\cdots,K\right\},\label{eqn:envelop}
%\end{IEEEeqnarray}
%where Lcov$\{\}$ denotes the $L$ value on the low envelope of the points inside the braces.
%Our main result is generalization of this result to  arbitrary $c,~r$.
%\end{remark}

\section{Main Result}\label{sec:result}

 Define
 \begin{IEEEeqnarray}{c}
 	c^*(r)\triangleq\frac{r}{K}+\left(1-\frac{r}{K}\right)\cdot g_r,\notag\\
 	L^*(r)\triangleq\frac{\lfloor r\rfloor+\lceil r\rceil-r}{\lfloor r\rfloor \lceil r\rceil }-\frac{1}{K},\notag
 \end{IEEEeqnarray}
 where
 \begin{IEEEeqnarray}{c}
 	g_r\triangleq \lfloor r\rfloor +\frac{(r-\lfloor r\rfloor)(K-\lceil r\rceil)}{K-r}.\notag
 \end{IEEEeqnarray}
 Notice that, $L^*(r)$ is the optimal \emph{storage-communication} tradeoff derived in \cite{Li2018Tradeoff}.

\begin{theorem}\label{thm:main} For any storage space $r\in[1,K)$,  and
\begin{IEEEeqnarray}{c}
c\in\bigg\{\frac{r}{K}+\left(1-\frac{r}{K}\right)g~:~g=1,\cdots,\lfloor r\rfloor\bigg\},\label{eqn:c_range}
\end{IEEEeqnarray}
  the optimal communication load $L^*(r,c)$ is given by
 \begin{IEEEeqnarray}{rCl}
 L^*(r,c)=\frac{1}{c-r/K}\cdot\left(1-\frac{r}{K}\right)^2.\label{eqn:optimalL}
 \end{IEEEeqnarray}
  For general $1\leq c\leq c^*(r)$, the optimal communication load $L^*(r,c)$ is given by the lower convex envelope of the points {in \eqref{eqn:c_range} and \eqref{eqn:optimalL}} and  the point $\left(c^*(r),L^*(r)\right)$.
  Moreover,
  \begin{IEEEeqnarray}{c}
  L^*(r,c)=L^*(r), \quad c^*(r)\leq c\leq r.\label{eqn:flat}
  \end{IEEEeqnarray}
\end{theorem}
\begin{IEEEproof}
	The tradeoff in Theorem~\ref{thm:main} is achieved by the D3C scheme, see \cite{Yan2018SCC}. 			Equality  \eqref{eqn:flat} has been shown in \cite[Corollary~1]{Yan2018SCC}.
%		Notice that, $L^*(r)$ is the optimal \emph{storage-communication} tradeoff \cite{Li2018Tradeoff}, i.e., the lower convex envelope of
%		\begin{IEEEeqnarray}{c}
%			L^*(r)=\frac{1}{r}\left(1-\frac{r}{K}\right),\quad r=1,\cdots,K.
%		\end{IEEEeqnarray}
The converse for the case $0\leq c\leq c^*(r)$ is proved in Section~\ref{sec:converse}.
	\end{IEEEproof}

\begin{figure}[!htb]
                \centering
               \includegraphics[width=0.49\textwidth,height=0.25\textwidth ]{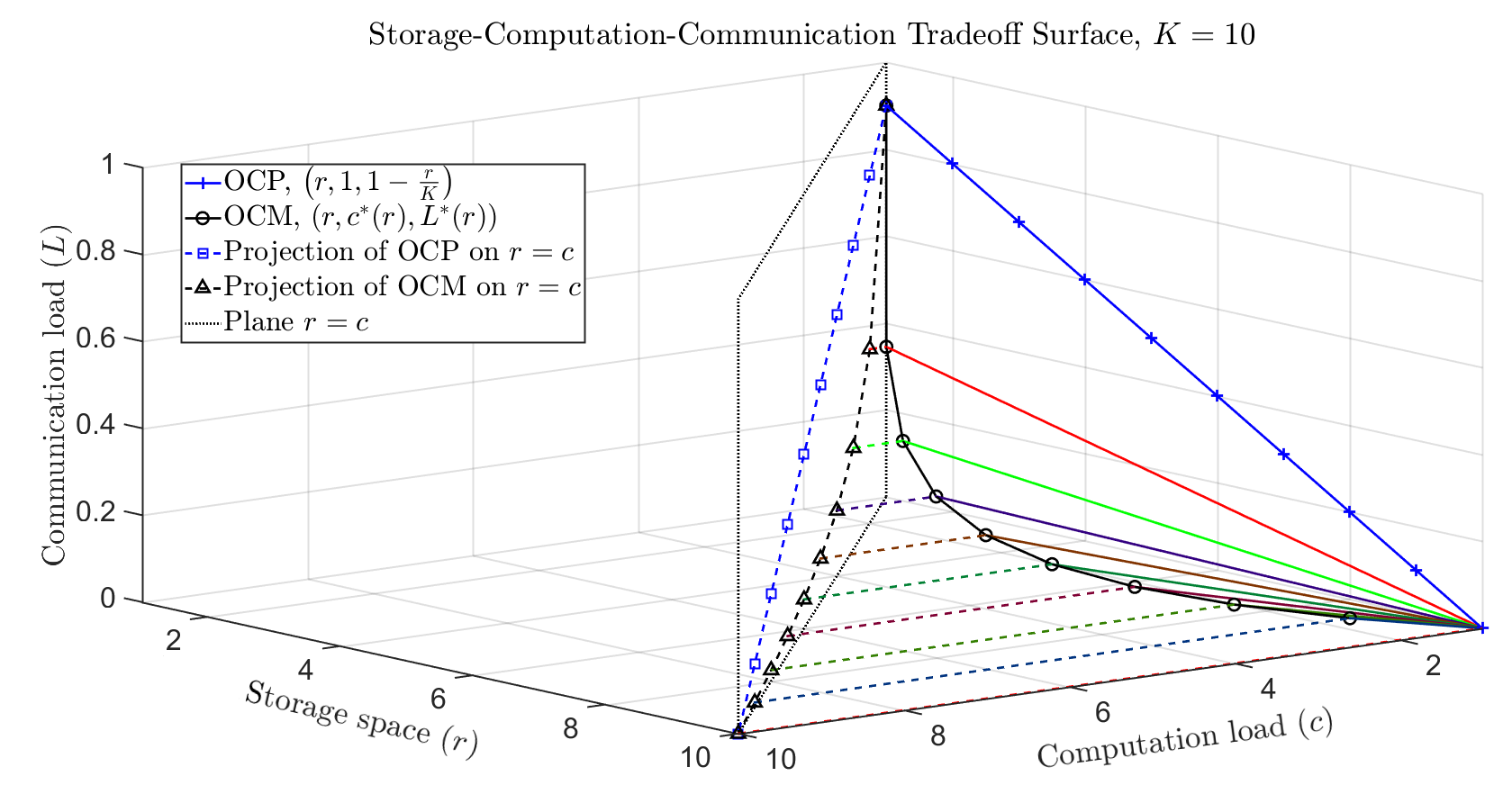}
                \caption{ The storage-computation-communication tradeoff surface for a system with $K=10$ nodes. The dashed lines are the projections of OCP and OCM to the plane $c=r$. }\label{Fig:tradeoff}
\end{figure}

Notice that,  $L^*(r,c)$ is piecewise linear in $(r,c)$.
In the \emph{storage-computation-communication (r-c-L)} space, where the
coordinates are associated with $r$, $c$, and $L$, respectively, Fig.~\ref{Fig:tradeoff} illustrates the  surface $L^*(r,c)$ characterized by Theorem \ref{thm:main} when $K=10$. In particular,
\begin{enumerate}
  \item The line
\begin{IEEEeqnarray}{c}
\left(r,1,1-\frac{r}{K}\right),\quad r\in[1,K)\notag
\end{IEEEeqnarray}
is the \emph{optimal computation curve (OCP)}, and characterizes the
optimal storage-communication tradeoff at the lowest computation load~($c=1$).
%in the sense that, for any $r$, it achieves the optimal computation load $c=1$ with the lowest communication load $1-r/K$;
  \item The curve
\begin{IEEEeqnarray}{c}
\left(r,c^*(r),L^*(r)\right),\quad r\in[1,K)\notag
\end{IEEEeqnarray}
is the \emph{optimal communication curve (OCM)}, and characterizes the
optimal storage-computation tradeoff at the lowest communication
load~($L=L^*(r)$).
%in the sense that, for any $r$, it achieves the optimal communication load $L^*(r)$ with the least computation load $c^*(r)$.
\item The pareto-optimal surface is given by the triangles between the OCP and OCM curves.
\end{enumerate}

\begin{remark} We  briefly sketch the D3C scheme in \cite{Yan2018SCC}, which achieves the optimal tradeoff in Theorem \ref{thm:main}. For integers $r,g$ such that $1\leq g\leq r<K$, the files are partitioned into ${K\choose r}{r\choose g}$ batches. Each batch is  associated with a tuple $(\mathcal{S},\mathcal{T})$ where $\mathcal{T}\subseteq \mathcal{S}\subset\mathcal{K}$, $|\mathcal{S}|=r,|\mathcal{T}|=g$. Let $\mathcal{W}_{\mathcal{S},\mathcal{T}}$ be the batch associated with $(\mathcal{S},\mathcal{T})$,  all  nodes in $\mathcal{S}$ store $\mathcal{W}_{\mathcal{S},\mathcal{T}}$, and compute their own  IVAs from $\mathcal{W}_{\mathcal{S},\mathcal{T}}$.   Only the nodes in $\mathcal{T}$ compute the IVAs from  $\mathcal{W}_{\mathcal{S},\mathcal{T}}$  that are  needed by the nodes in $\mathcal{K}\backslash\mathcal{S}$.  In the shuffle phase, for each pair $(\mathcal{I},\mathcal{J})$ such that $\mathcal{I}=r+1,\mathcal{J}=g+1$, each node $k$ in $\mathcal{J}$ creates a coded multicast signal useful for all nodes in $\mathcal{J}\backslash\{k\}$. Based on the received multicast signals and the IVAs it  computed locally, each node can then compute the desired output function in the reduce phase.

When $g=r$, the D3C degrades to the modified CDC (M-CDC) scheme in \cite{Fragouli}. The M-CDC scheme achieves the $K$ corner points of the optimal tradeoff surface.  Time- and memory- sharing the M-CDC scheme with different parameters can thus achieve all pareto-optimal points on the tradeoff surface, see \cite{Yan2018SCC} and \cite{Yan2018SCCfull} for details.

One may observe that, in both the D3C  and the M-CDC scheme, the required number of input files  increases very fast with the number of nodes. This may prevent  implementation in practice. In the longer version of this paper \cite{Yan2018SCCfull}, we  propose ways to decrease the required number of files via placement delivery arrays \cite{Yan2017PDA}.
\end{remark}

%{\color{red}The projections of OCP and OCM on the plane $r=c$ (plotted in
%  Fig. \ref{Fig:tradeoff}) can be interpreted as the tradeoffs achieved by the traditional uncoded scheme
%  and by the coded distributed computation scheme in %\emph{optimal storage-communication tradeoff} investigated in
%  \cite{Li2018Tradeoff}. In this sense, the uncoded scheme achieves the
%  optimal computation load, and Theorem \ref{thm:main} establishes the
%  transition from the optimal computation load to the optimal communication load for any given $r$.}
%
%
%

\section{Converse}\label{sec:converse}

 Fix $r\in[1,K)$, and $c\in[1,r]$. Consider a file allocation $\mathcal{M}=\{\mathcal{M}_k\}_{k=1}^K$ and its feasible IVA sets $\mathcal{C}=\{\mathcal{C}_{k}\}_{k=1}^K$, so that\footnote{As  $\epsilon>0$ can be arbitrarily close to $0$ in Definition \ref{definition:ach}, to derive the lower bound for $L^*(r,c)$, we need to consider the case $\epsilon\rightarrow 0$. }
\begin{IEEEeqnarray}{rCl}
\frac{\sum_{k=1}^K|\mathcal{M}_k|}{N}&\leq&r,\label{const:M}\\
\frac{\sum_{k=1}^K|\mathcal{C}_k|}{NK}&\leq&c.\label{const:C}
\end{IEEEeqnarray}
For any nonempty set $\mathcal{S}\subseteq[K]$, denote
$
X_{\mathcal{S}}\triangleq\{X_k\}_{k\in\mathcal{S}},
\mathcal{V}_{\mathcal{S}}\triangleq \cup_{k\in\mathcal{S}}\mathcal{V}_k,
\mathcal{C}_{\mathcal{S}}\triangleq \cup_{k\in\mathcal{S}}\mathcal{C}_k
$.
 For any $k\in\mathcal{S}$ and $j\in[|\mathcal{S}|-1]$, define
\begin{IEEEeqnarray}{rCl}
\mathcal{B}_{\mathcal{S},j}^{k}&\triangleq&\{v_{k,n}:v_{k,n}~\mbox{is only computed}\notag\\
 &&\qquad\qquad\qquad\mbox{by $j$ nodes in}~ \mathcal{S}\backslash\{k\}\}.\IEEEeqnarraynumspace \notag
\end{IEEEeqnarray}
Let  $b_{\mathcal{S},j}^k$ be the cardinality of $\mathcal{B}_{\mathcal{S},j}^{k}$. Then the cardinality of
\begin{IEEEeqnarray}{c}
\mathcal{B}_{\mathcal{S},j} \triangleq
\bigcup_{k\in\mathcal{S}}\mathcal{B}_{\mathcal{S},j}^{k}\notag
\end{IEEEeqnarray}
 is given by
\begin{IEEEeqnarray}{c}
b_{\mathcal{S},j}\triangleq\sum_{k\in\mathcal{S}}b_{\mathcal{S},j}^k.\label{eqn:bsjk}
\end{IEEEeqnarray}

\subsection{Auxiliary Lemmas}

To prove the converse, we need the following two lemmas, where   Lemma \ref{Lemma:2} is proved  in Section~\ref{sec:proof_lemma}.

\begin{lemma}\label{Lemma:2} For any nonempty set $\mathcal{S}\subseteq[K]$,
\begin{IEEEeqnarray}{c}
H(X_{\mathcal{S}}|\mathcal{V}_{\mathcal{S}^c},\mathcal{C}_{\mathcal{S}^c})\geq T\sum_{j=1}^{|\mathcal{S}|-1}b_{\mathcal{S},j}\cdot\frac{1}{j}, \label{eqn:lemma2}
\end{IEEEeqnarray}
where $\mathcal{S}^c\triangleq [K]\backslash\mathcal{S}$.
\end{lemma}

\begin{lemma}\label{Lemma:3}  Consider set $\mathcal{S}=[K]$ and define % \eqref{eqn:Bj} and \eqref{eqn:bsjk}, let $\mathcal{S}=[K]$ and
	$b_j\triangleq b_{[K],j}$. Then,
\begin{IEEEeqnarray}{rCl}
\sum_{j=1}^{K-1}b_j&\geq& N(K-r),\label{eqn:ineq1}\\
\sum_{j=1}^{K-1}(j-1)b_j&\leq&(c-1)NK.\label{eqn:ineq2}
\end{IEEEeqnarray}
\end{lemma}

\begin{IEEEproof} For any $k\in[K]$, define
\begin{IEEEeqnarray}{l}
\mathcal{A}_k\triangleq\left\{v_{k,n}:v_{k,n}~\mbox{is computed by
node}~k,n\in[N]\right\}.\IEEEeqnarraynumspace\notag
\end{IEEEeqnarray}
Set $a_k=|\mathcal{A}_k|$. Notice that
\begin{IEEEeqnarray}{c}
\mathcal{A}_1,\cdots,\mathcal{A}_K,\mathcal{B}_{[K],1},\cdots,\mathcal{B}_{[K],K-1}\notag
\end{IEEEeqnarray}
form a partition of all IVAs, and therefore
\begin{IEEEeqnarray}{c}
\sum_{k=1}^Ka_k+\sum_{j=1}^{K-1}b_j=NK.\label{eqn:sumeq}
\end{IEEEeqnarray}
Moreover, since node $k$ must store $w_n$ if it has computed $v_{k,n}$, it must hold that $a_k\leq|\mathcal{M}_k|,$ and thus by \eqref{const:M},
\begin{IEEEeqnarray}{c}
\sum_{k=1}^Ka_k\leq\sum_{k=1}^{K}|\mathcal{M}_k|\leq rN.\label{eqn:sumineq1}
\end{IEEEeqnarray}
Finally, for each $k\in[K],\, j\in[K-1]$, the IVAs in $\mathcal{A}_k$ must be computed at node $k$ and IVAs $\mathcal{B}_{[K],j}$ must be computed at $j$ nodes, and by \eqref{const:C},
\begin{IEEEeqnarray}{c}
\sum_{k=1}^Ka_k+\sum_{j=1}^{K-1}jb_j\leq\sum_{k=1}^K|\mathcal{C}_k|\leq cNK.\label{eqn:sumineq2}
\end{IEEEeqnarray}
From \eqref{eqn:sumeq}--\eqref{eqn:sumineq2}, we obtain \eqref{eqn:ineq1} and \eqref{eqn:ineq2}.
\end{IEEEproof}

\subsection{Proof of the Converse to Theorem \ref{thm:main}}

 %We first derive a lower bound on the communication load $L$.
%Applying Lemma \ref{Lemma:2} to $\mathcal{S}=[K]$, we obtain:
%\begin{IEEEeqnarray}{rCl}
%L&\geq& \frac{H\left(X_{[K]}\right)}{NKT}\\
%&\geq&\sum_{j=1}^{K-1}\frac{b_j}{NK}\cdot\frac{1}{j}\label{ineq:lowerbound1}\\
%&=&\frac{\sum_{l=1}^{K-1}b_l}{NK}\cdot\sum_{j=1}^{K-1}\frac{b_j}{\sum_{l=1}^{K-1}b_l}\cdot\frac{1}{(j-1)+1}\\
%&\overset{(a)}{\geq}&\frac{\sum_{l=1}^{K-1}b_l}{NK}\cdot\frac{1}{\frac{\sum_{j=1}^{K-1}(j-1)b_j}{\sum_{l=1}^{{{K-1}}}b_l}+1}\\
%&\overset{(b)}{\geq}&\frac{\sum_{l=1}^{K-1}b_l}{NK}\cdot\frac{1}{\frac{(c-1)NK}{\sum_{l=1}^{{K-1}}b_l}+1}\\
%&\overset{(c)}\geq&\frac{N(K-r)}{NK}\cdot\frac{1}{\frac{(c-1)NK}{N(K-r)}+1}\\
%&=&\frac{1}{c-r/K}\cdot\left(1-\frac{r}{K}\right)^2,\label{bound1}
%\end{IEEEeqnarray}
%where $(a)$ follows from the fact that the function $\frac{1}{x+1}$ is convex over $x\in\mathbb{R}^+$ and Jensen's inequality, $(b)$ follows from \eqref{eqn:ineq2}, and
%$(c)$ follows from \eqref{eqn:ineq1} and the fact that the function
%$
%\frac{x}{NK}\cdot\frac{1}{{(c-1)NK}/{x}+1}
%$
%increases with $x$ over $x\in\mathbb{R}^+$.
%{This proves the converse for  $r \in [1,K)$ and $c$ belonging to the set \eqref{eqn:c_range}.}
%

For each $c\in[1,
r]$, define
\begin{IEEEeqnarray}{c}
g\triangleq\frac{c-r/K}{1-r/K}.\notag
\end{IEEEeqnarray}
Notice that $g\geq 1$ since we assume $c\geq 1$.  Let $g_1\triangleq\lfloor g\rfloor,$
$g_2\triangleq \lceil g\rceil$, and
\begin{IEEEeqnarray}{rCl}
c_1&=&\frac{r}{K}+\left(1-\frac{r}{K}\right)g_1,\label{eqn:c1}\\
c_2&=&\frac{r}{K}+\left(1-\frac{r}{K}\right)g_2.\label{eqn:c2}
\end{IEEEeqnarray}
{Notice that by these definitions,
\begin{equation}
c_1 \leq c \leq c_2.
\end{equation}}

Choose $\lambda,\mu\in\mathbb{R}$ so that
\begin{IEEEeqnarray}{rCl}
\lambda x+\mu|_{x=c_1}&=&\frac{1}{c_1-r/K}\cdot\left(1-\frac{r}{K}\right)^2,\label{eqn:x1}\\
%&=&\frac{1}{g_1}\big(1-\frac{r}{K}\big),\\
\lambda x+\mu|_{x=c_2}&=&\frac{1}{c_2-r/K}\cdot\left(1-\frac{r}{K}\right)^2.\label{eqn:x2}
%&=&\frac{1}{g_2}\big(1-\frac{r}{K}\big).
\end{IEEEeqnarray}
Then from \eqref{eqn:c1}--\eqref{eqn:x2}, and the fact $g_2-g_1=1$,  we conclude that $\lambda$ and $\mu$ satisfy:
\begin{IEEEeqnarray}{rCl}
\lambda&=&\frac{1}{g_2}-\frac{1}{g_1}<0,\label{eqn:lambda}\\
\mu&=&\frac{c_2}{g_1}-\frac{c_1}{g_2}>0,\notag\\
\lambda+\mu&=&\frac{c_2-1}{g_1}-\frac{c_1-1}{g_2}>0. \label{eqn:lambdamu}
\end{IEEEeqnarray}
By the convexity of the function $\frac{1}{x-r/K}\big(1-\frac{r}{K}\big)^2$ over $x\in[1,+\infty)$, we then obtain:
\begin{IEEEeqnarray}{l}
\frac{1}{x-r/K}\left(1-\frac{r}{K}\right)^2\geq \lambda x+\mu,\notag\\
~\forall~x\in\left\{\frac{r}{K}+\left(1-\frac{r}{K}\right)g:g=1,\cdots,K-1\right\}.\notag
\end{IEEEeqnarray}
Therefore,
\begin{IEEEeqnarray}{rCl}
L&\geq& \frac{H\left(X_{[K]}\right)}{NKT}\notag\\
&\geq&\sum_{j=1}^{K-1}\frac{b_j}{NK}\cdot\frac{1}{j}\notag\\
&\geq& \frac{1}{N(K-r)}\notag
\sum_{j=1}^{K-1}\frac{b_j}{\left(1-\frac{r}{K}\right)j+\frac{r}{K}-\frac{r}{K}}\left(1-\frac{r}{K}\right)^2\IEEEeqnarraynumspace\notag\\
&\geq&\frac{1}{N(K-r)}\sum_{j=1}^{K-1}b_j\left(\lambda \left(\left(1-\frac{r}{K}\right)j+\frac{r}{K}\right)+\mu\right)\notag\\
%&=&\frac{1}{N(K-r)}\cdot\left(\lambda\big(1-\frac{r}{K}\big)\cdot\sum_{j=1}^{K-1}jb_j
%+\big(\lambda\frac{r}{K}+\mu\big)\cdot\sum_{j=1}^{K-1}b_j\right)\\
&=&\frac{\lambda}{NK}\cdot\sum_{j=1}^{K-1}(j-1)b_j+\frac{\lambda+\mu}{N(K-r)}\cdot\sum_{j=1}^{K-1}b_j\notag\\
&\overset{(a)}{\geq}&\frac{\lambda}{NK}\cdot (c-1)NK+\frac{\lambda+\mu}{N(K-r)}\cdot N(K-r)\notag\\
&=&\lambda c+\mu,\label{bound2}
\end{IEEEeqnarray}
where $(a)$ follows from \eqref{eqn:ineq1}, \eqref{eqn:ineq2},
\eqref{eqn:lambda} and \eqref{eqn:lambdamu}. {This implies that for any storage space $r\in[1,K)$ and computation load $c\in[c_1,c_2]$, the optimal communication load  $L^*(r,c)$ is lower bounded by the lower convex envelope of $(c_1,L^*(r,c_1))$ and $(c_2,L^*(r,c_2))$. Noting that also the
point $\left(c^*(r),L^*(r)\right)$ is on the line \eqref{bound2} concludes the converse proof. }%. Thus, for any given $\mathcal{M}$ and $\mathcal{C}$ satisfying \eqref{const:M} and \eqref{const:C}, the communication load $L$ is lower bounded by the $L$-value of the lower convex envelope formed by the points in \eqref{eqn:c_range}, \eqref{eqn:optimalL} and the point $\left(c^*(r),L^*(r)\right)$. Furthermore, as it works for all $\mathcal{M},\mathcal{C}$  subject to \eqref{const:M} and \eqref{const:C} and
%\begin{IEEEeqnarray}{rCl}
% L^*(r,c)&=&\inf\left\{L_{\mathcal{M},\mathcal{C}}:\mathcal{M}, \mathcal{C}~\mbox{satisfies \eqref{const:M} and \eqref{const:C}}\right\},\IEEEeqnarraynumspace
% \end{IEEEeqnarray}
%the optimal communication load $L^*(r,c)$ is lower bounded by the same value. As a result,
% $L^*(r,c)$ is lower bounded by the lower convex envelope formed by the points in \eqref{eqn:c_range} and \eqref{eqn:optimalL} for any  $1\leq c\leq c^*(r)$. } %Finally,  as $L^*(c,r)$ is non-increasing with $c$ for fixed $r$,
%\begin{IEEEeqnarray}{c}
%L^*(c,r)\geq L^*(r,r), \quad \forall~c^*(r)\leq c\leq r.
%\end{IEEEeqnarray}

\subsection{Proof of Lemma \ref{Lemma:2}}\label{sec:proof_lemma}

For notational brevity, we denote the tuple
$(\mathcal{V}_\mathcal{S},\mathcal{C}_{\mathcal{S}})$ by
$Y_{\mathcal{S}}$ for any $\mathcal{S}\subseteq[K]$.
We prove Lemma \ref{Lemma:2} by mathematical induction on the size of $\mathcal{S}$:

When $|\mathcal{S}|=1$, without loss of generality, assume $\mathcal{S}=\{k\}$, then \eqref{eqn:lemma2} becomes
$
H\left(X_k|Y_{[K]\backslash\{k\}}\right)\geq 0,
$
which is trivial.

Suppose that, the statement is true for all subsets of $[K]$ with size
$s$, $1\leq s<K$. Consider a set $\mathcal{S}\subseteq[K]$ such that $|\mathcal{S}|=s+1$. Then
\begin{align}
%\begin{IEEEeqnarray}{rCl}
  \MoveEqLeft{H(X_{\mathcal{S}}|Y_{\mathcal{S}^c})}\nonumber\\
&=\frac{1}{|\mathcal{S}|}\sum_{k\in\mathcal{S}}H(X_{\mathcal{S}},X_k|Y_{\mathcal{S}^c})\notag\\
&=\frac{1}{|\mathcal{S}|}\sum_{k\in\mathcal{S}}\big(H(X_k|Y_{\mathcal{S}^c})+H(X_{\mathcal{S}}|X_k,Y_{\mathcal{S}^c})\big)\notag\\
%&=&\frac{1}{|\mathcal{S}|}\sum_{k\in\mathcal{S}}H(X_k|Y_{\mathcal{S}^c})+\frac{1}{|\mathcal{S}|}\sum_{k\in\mathcal{S}} H(X_\mathcal{S}|X_k,Y_{\mathcal{S}^c})\\
&\geq\frac{1}{|\mathcal{S}|} H(X_{\mathcal{S}}|Y_{\mathcal{S}^c})+\frac{1}{|\mathcal{S}|}\sum_{k\in\mathcal{S}} H(X_\mathcal{S}|X_k,Y_{\mathcal{S}^c}).\label{eqn:step1}
%\end{IEEEeqnarray}
\end{align}
   Then from \eqref{eqn:step1}, we have
  \begin{align}%{rCl}
    \MoveEqLeft[0]{H(X_{\mathcal{S}}|Y_{\mathcal{S}^c})} \nonumber \\
  &\geq\frac{1}{|\mathcal{S}|-1}\sum_{k\in\mathcal{S}} H(X_\mathcal{S}|X_k,Y_{\mathcal{S}^c})\nonumber\\
  &\geq\frac{1}{s}\sum_{k\in\mathcal{S}}H(X_\mathcal{S}|X_k,\mathcal{C}_k, Y_{\mathcal{S}^c})\notag\\
  &\overset{(a)}{=} \frac{1}{s}\sum_{k\in\mathcal{S}}H(X_{\mathcal{S}}|\mathcal{C}_k,Y_{\mathcal{S}^c})\notag\\
  &\overset{(b)}{=} \frac{1}{s}\sum_{k\in\mathcal{S}}\big(H(X_{\mathcal{S}}|\mathcal{C}_k,Y_{\mathcal{S}^c})+H(\mathcal{V}_k|X_{\mathcal{S}},\mathcal{C}_k,Y_{\mathcal{S}^c})\big)\notag\\
&\overset{(c)}{=} \frac{1}{s}\sum_{k\in\mathcal{S}}H(X_{\mathcal{S}},\mathcal{V}_k|\mathcal{C}_k,Y_{\mathcal{S}^c})\notag\\
&\overset{(d)}{=} \frac{1}{s}\sum_{k\in\mathcal{S}}\left(H(\mathcal{V}_k|\mathcal{C}_k,Y_{\mathcal{S}^c})+H(X_{\mathcal{S}}|\mathcal{V}_k,\mathcal{C}_k,Y_{\mathcal{S}^c})\right)\notag\\
%&=&\frac{1}{s}\sum_{k\in\mathcal{S}}\big(H(\mathcal{V}_k|\mathcal{C}_k,\mathcal{V}_{\mathcal{S}^c},\mathcal{C}_{\mathcal{S}^c})\notag\\
%&&\qquad\qquad+H(X_{\mathcal{S}\backslash\{k\}},X_k|\mathcal{V}_k,\mathcal{C}_k,Y_{\mathcal{S}^c})\big)\\
&\overset{(e)}{=}
\frac{1}{s}\sum_{k\in\mathcal{S}}\left(H(\mathcal{V}_k|\mathcal{C}_{(\mathcal{S}\backslash\{k\})^c})+H(X_{\mathcal{S}\backslash\{k\}}|Y_{(\mathcal{S}\backslash\{k\})^c})\right)
%\IEEEeqnarraynumspace
\notag\\
&\overset{(f)}{\geq} \frac{T}{s}\sum_{k\in\mathcal{S}}\sum_{j=1}^{s}b_{\mathcal{S},j}^k+\frac{T}{s}\sum_{k\in\mathcal{S}}\sum_{j=1}^{s-1}b_{\mathcal{S}\backslash\{k\},j}\cdot\frac{1}{j}\notag\\
&= \frac{T}{s}\sum_{j=1}^{s}\sum_{k\in\mathcal{S}}b_{\mathcal{S},j}^k+\frac{T}{s}\sum_{j=1}^{s-1}\sum_{k\in\mathcal{S}}b_{\mathcal{S}\backslash\{k\},j}\cdot\frac{1}{j}\notag\\
&\overset{(g)}{=}\frac{T}{s}\sum_{j=1}^{s}b_{\mathcal{S},j}+\frac{T}{s}\sum_{j=1}^{s-1}\sum_{k\in\mathcal{S}}\sum_{l\in\mathcal{S}\backslash\{k\}}b_{\mathcal{S}\backslash\{k\},j}^l\cdot\frac{1}{j}\notag\\
%&=&\frac{T}{s}\sum_{j=1}^{s}b_{\mathcal{S}}^j+\frac{T}{s}\sum_{j=1}^{s-1}\sum_{k\in\mathcal{S}}\sum_{l\in\mathcal{S}\backslash\{k\}} b_{\mathcal{S}\backslash\{k\},l}^j\cdot\frac{1}{j},\\
&=\frac{T}{s}\sum_{j=1}^{s}b_{\mathcal{S},j}+\frac{T}{s}\sum_{j=1}^{s-1}\sum_{l\in\mathcal{S}}\sum_{k\in\mathcal{S}\backslash\{l\}} b_{\mathcal{S}\backslash\{k\},j}^l\cdot\frac{1}{j},\label{eqn:continue}
%&\overset{(b)}{=}&\frac{T}{s}\sum_{j=1}^{s}b_{\mathcal{S}}^j+\frac{T}{s}\sum_{j=1}^{s-1}\sum_{l\in\mathcal{S}} b_{\mathcal{S},l}^{j}\cdot\frac{s-j}{j},\\
%&=&\frac{T}{s}\cdot\sum_{j=1}^{s}b_{\mathcal{S}}^j+\frac{T}{s}\cdot\sum_{j=1}^{s-1}b_{\mathcal{S}}^j\cdot\frac{s-j}{j}\\
%%&=&\frac{T}{s}\cdot\sum_{j=1}^{s}b_{\mathcal{S}}^j+T\cdot\sum_{j=1}^{s-1}\frac{b_{\mathcal{S}}^j}{j}-\frac{T}{s}\cdot\sum_{j=1}^{s-1}b_{\mathcal{S}}^j\\
%%&=&T\cdot\frac{b_{\mathcal{S}}^s}{s}+T\cdot\sum_{j=1}^{s-1}\frac{b_{\mathcal{S}}^j}{j}\\
%&=&T\cdot\sum_{j=1}^{|\mathcal{S}|-1}\frac{b_{\mathcal{S}}^j}{j}.
\end{align}
  %\end{IEEEeqnarray}
where  $(a)$ holds because $X_k$ is a function of $\mathcal{C}_k$; $(b)$ holds because by
$
H(\mathcal{V}_k|X_{\mathcal{S}},\mathcal{C}_k,Y_{\mathcal{S}^c})=0,
$
 since $\mathcal{V}_k$ can be decoded using
 $\mathcal{C}_k,X_{\mathcal{S}}$ and $X_{\mathcal{S}^c}$, which is a
 function of $Y_{\mathcal{S}^c}$; $(c)$ and $(d)$ follow from the chain rule;  $(e)$ holds because $Y_{\mathcal{S}^c}=(\mathcal{V}_{\mathcal{S}^c},\mathcal{C}_{\mathcal{S}^c})$ and by the independence between $\mathcal{V}_k$ and $\mathcal{V}_{\mathcal{S}^c}$; $(f)$ holds by the definition of $b_{\mathcal{S},j}^k$ and the induction assumption; and $(g)$ holds by \eqref{eqn:bsjk}.

 Notice that, in \eqref{eqn:continue},
\begin{IEEEeqnarray}{rCl}
&&\sum_{k\in\mathcal{S}\backslash\{l\}} b_{\mathcal{S}\backslash\{k\},j}^l\notag\\
%&\overset{(a)}{=}&\sum_{l\in\mathcal{S}}\sum_{k\in\mathcal{S}\backslash\{l\}} \sum_{n=1}^N\mathbb{I}(v_{l,n}~\mbox{is only  }
%\mbox{computed by $j$ nodes in }\mathcal{S}\backslash\{k,l\})\\
&\overset{(a)}{=}&\sum_{k\in\mathcal{S}\backslash\{l\}} \sum_{n=1}^N\mathbb{I}(v_{l,n}~\mbox{is only computed  by $j$ nodes  in} \notag\\
&&\qquad\mathcal{S}\backslash\{l\})\cdot\mathbb{I}(v_{l,n}~\mbox{is not computed by node}~k)\IEEEeqnarraynumspace\notag\\
&=&\sum_{n=1}^N\mathbb{I}(v_{l,n}~\mbox{is only computed  by $j$ nodes in }\mathcal{S}\backslash\{l\}) \notag\\
&&\qquad\cdot\sum_{k\in\mathcal{S}\backslash\{l\}} \mathbb{I}(v_{l,n}~\mbox{is not computed by node}~k)\notag\\
&=&\sum_{n=1}^N\mathbb{I}(v_{l,n}~\mbox{is only computed by $j$ nodes in } \mathcal{S}\backslash\{l\})\notag\\
&&\qquad\cdot~(s-j)\notag\\
&\overset{(b)}{=}&b_{\mathcal{S},j}^l(s-j), \notag
\end{IEEEeqnarray}
where $(a)$ and $(b)$ follow from the definition of  $b_{\mathcal{S},j}^l$. Thus, with \eqref{eqn:continue},
\begin{IEEEeqnarray}{rCl}
H(X_{\mathcal{S}}|Y_{\mathcal{S}^c})
&\geq&\frac{T}{s}\sum_{j=1}^{s}b_{\mathcal{S},j}+\frac{T}{s}\sum_{j=1}^{s-1}\sum_{l\in\mathcal{S}} b_{\mathcal{S},j}^{l}\cdot\frac{s-j}{j}\IEEEeqnarraynumspace\notag\\
&\overset{(a)}{=}&\frac{T}{s}\sum_{j=1}^{s}b_{\mathcal{S},j}+\frac{T}{s}\sum_{j=1}^{s-1}b_{\mathcal{S},j}\cdot\frac{s-j}{j}\notag\\
%&=&\frac{T}{s}\cdot\sum_{j=1}^{s}b_{\mathcal{S}}^j+T\cdot\sum_{j=1}^{s-1}\frac{b_{\mathcal{S}}^j}{j}-\frac{T}{s}\cdot\sum_{j=1}^{s-1}b_{\mathcal{S}}^j\\
%&=&T\cdot\frac{b_{\mathcal{S}}^s}{s}+T\cdot\sum_{j=1}^{s-1}\frac{b_{\mathcal{S}}^j}{j}\\
&=&T\sum_{j=1}^{|\mathcal{S}|-1}\frac{b_{\mathcal{S},j}}{j}.\notag
  \end{IEEEeqnarray}
  where $(a)$ follows from \eqref{eqn:bsjk}.

Notice that, we have proved that \eqref{eqn:lemma2} holds for all
$\mathcal{S}\subseteq[K]$ with $|\mathcal{S}|=s+1$. By  induction, we conclude that \eqref{eqn:lemma2} holds for
all nonempty subsets $\mathcal{S}\subseteq[K]$.

\section{Conlusion}\label{sec:conclusion}

We proved a converse matching the performance of our recently proposed D3C \cite{Yan2018SCC}.  As a result, the pareto-optimal storage-computation-communication  tradeoff surface of all achievable storage-computation-communication triples is characterized.

\section*{Acknowledgement}
The work of Q. Yan and M. Wigger has been supported by the ERC under grant agreement 715111.

\ifCLASSOPTIONcaptionsoff
  \newpage
\fi

\end{document}